\documentclass[titlepage,11pt,a4paper,reqno]{article}
\usepackage{amsmath,amssymb,amsfonts,graphics,setspace,tensor}
\usepackage{slashed}
\usepackage[linktocpage=true,colorlinks,pdftex]{hyperref}
\usepackage{xcolor}
\colorlet{linkequation}{blue}
\usepackage{bm}
\usepackage{doi}
\usepackage{hyperref}
\hypersetup{colorlinks, citecolor=violet, filecolor=black, linkcolor=black, urlcolor=blue}

\allowdisplaybreaks
\ifx\HCode\UnDef\else\hypersetup{tex4ht}\fi 

\begin{document}


\begin{titlepage}

\centerline{\Large \bf Fractional Mimetic Dark Matter Model \bf}
\centerline{\Large \bf in Fractional Action-like Variational Approach \bf} 
\vskip 1cm
\centerline{\bf C. F. L. Godinho and I. V. Vancea}
\vskip 0.5cm
\centerline{\sl Group of Theoretical Physics and Mathematical Physics,
Department of Physics}
\centerline{\sl Federal Rural University of Rio de Janeiro}
\centerline{\sl Cx. Postal 23851, BR 465 Km 7, 23890-000 Serop\'{e}dica - RJ,
Brazil}
\centerline{
\texttt{\small crgodinho@gmail.com; ionvancea@ufrrj.br} 
}
\vspace{0.5cm}

\centerline{27 August 2022}

\vskip 1.4cm
\centerline{\large\bf Abstract}
In this paper, we propose a new model of fractional mimetic dark matter based on the fractional action-like variational approach (FALVA) implementation of fractionality. The model is non-local at classical level and its equations of motion are highly non-linear. We discuss two particular cases of non-fractional and fractional mimetic dark fields and 
solve completely the equations of motion of these models. In particular, we obtain the metric of the fractional physical space-time. Also, we obtain the basic geometrical objects in the two cases.

\vskip 0.7cm 

\noindent

\end{titlepage}


\section{Introduction}

The substitution of the standard mathematical formalism used to construct classical gravitational models, that is differential geometry based on the standard calculus in flat space-time, by differential objects formulated in terms of fractional calculus and FALVA formalism to the action functional
\cite{Nabulsi:2008a} - \cite{Nabulsi:2009b}, introduces new phenomena such as nonlocality, memory effects, etc. The generalization of the Einstein-Hilbert action and the construction of the fractional differential geometry, including  in FALVA formulation was given in \cite{El-Nabulsi:2009bup} - \cite{El-Nabulsi:2021uwp}. The the fundamental covariant action of fractional gravity was given for the first time in \cite{Calcagni:2021aap}.
These approaches have been used to study cosmological models in \cite{El-Nabulsi:2007wgc}-\cite{Shchigolev:2015rei}. Different aspects of the fractional Newtonian gravity have been discussed in
\cite{Giusti:2020rul}-\cite{Landim:2021www} and the gravitons obtained from the FALVA model have been examined in \cite{El-Nabulsi:2012wpc}. Classical models of gravity with nonlocal and memory effects have been proposed in \cite{Mashhoon:2006mj,Chicone:2015sda}, too, but there the nonlocality was the result of the generalization to gravity of the observation that the measurements in field theory are actually averages over a time interval (see also the monograph \cite{Mashhoon:2017qyw}).

In this paper, we propose a new fractional nonlocal gravitational model
that generalizes the ones discussed in the literature and mentioned above by including a dark matter component. Our proposal is based on the mimetic dark matter model firstly put forward in \cite{Chamseddine:2013kea} in the Lagrangian formulation developed in \cite{Golovnev:2013jxa}-\cite{Malaeb:2014vua} and in the general cosmological background discussed in \cite{Chamseddine:2014vna,Chamseddine:2016uef} and \cite{Cognola:2016gjy}-\cite{Vagnozzi:2017ilo}). What sets apart the mimetic gravity from other modified theories of gravity is the fact that the dark matter component is a purely geometrical effect with no matter fields associated to it. In the above setting, the fields of the model are spatially homogeneous which makes the model suitable to generalization by the FALVA method. Fractional cosmological models with this properties in the FALVA approach have been constructed in \cite{Shchigolev:2010vh,Shchigolev:2021lbm}. Therefore, the fractioanl mimetic dark matter model (FMDDM) proposed here is a direct generalization of the models from \cite{Shchigolev:2010vh,Shchigolev:2021lbm} (see also
for applications of fractional calculus to quantum cosmology
\cite{Moniz:2020emn}-\cite{Jalalzadeh:2022uhl} and
\cite{Calcagni:2021mmj} for an interesting fractional dark matter model).

Since the nonlocality is introduced by the mathematical formalism, it is natural to ask what kind of phenomenology FMDDM could be representing. As discussed above, the nonlocality in the time variable realizes in a new way the memory effects from \cite{Mashhoon:2017qyw}. Also, since the FALVA formalism contains terms that can be interpreted, at least in some fractional mechanical systems, as nonconservative forces, the FMDDM realizes a new type of nonconservative gravity \cite{Mignani:1997a,Fabris:2017msx} 
In order to implement the fractionality, we use the formulation of the FMDDM in terms of the degrees of freedom of the auxiliary metric. From that, it follows that the physical space-time geometry inherits fractionality properties which are reflected in phenomenological parameters that can be used to test fractional structure. 

The paper is organized as follows. In Section 2, we briefly review the mimetic dark matter model. In Section 3, we present the FMDMM in the FALVA approach. Here, we give the action and derive the Euler-Lagrange equations. In this section, we also solve the Euler-Equations for a background characterized by a constant auxiliary Hubble parameter. We obtain the general solution parametrized by the fractionality parameter. In Section 4, we calculate the physical Ricci tensor corresponding to the general solution and discuss the cosmological redshift and the cosmological distances in two limits for different values of the fractionality parameter. We drive our conclusions in the last section. Throughout this work, we use the space-time metric with signature $(-, +, +, + )$ and the natural units with $c = 1$. 

\section{Mimetic Dark Matter}

In this section, we briefly review the formulation of the mimetic dark matter model. The results presented here can be found in the original works \cite{Golovnev:2013jxa}-\cite{Malaeb:2014vua} which we refer to for further details.

The mimetic gravity is a generalization of the General Relativity in which the physical space-time metric $g_{\mu \nu}$ is parametrized in terms of the conformal scalar degree of freedom $\phi$, also called mimetic field.
This is realized by introducing an auxiliary metric $l_{\mu \nu}$ which is related to $g_{\mu \nu}$ and $\phi$ by the following equation
\begin{equation}
g_{\mu\nu} 
:= 
\left( 
l^{\alpha \beta} \partial_\alpha \phi \partial_\beta \phi 
\right) l_{\mu\nu}
=
\Phi^2  l_{\mu\nu}
\, .
\label{MDM-metric}
\end{equation}
In this setting, $g_{\mu \nu}$ is manifestly invariant under the 
Weyl transformations of the auxiliary metric
\begin{equation}
l_{\mu \nu} \to {l'}_{\mu \nu} = \omega^{2} l_{\mu \nu}
\, ,
\label{Auxiliary-metric-Weyl}
\end{equation} 
where $\omega = \omega (x)$ is an arbitrary function on the coordinates. 
A straightforward consistency condition follows from the above equation
\begin{equation}
\Phi^2
=
l^{\mu \nu}
\partial_{\mu} \phi \partial_{\nu} \phi
\, .
\label{MDM-consistency}
\end{equation}
The dynamics of the mimetic gravity is governed by the Einstein-Hilbert action expressed in terms of the field variables $l_{\mu \nu}$ and $\phi$
\begin{equation}
S_0 [l_{\mu\nu},\phi]  = 
- \frac{1}{2 \kappa}\int d^4 x \sqrt{-g ( l_{\mu\nu},\phi ) }
\,
R \left[ g_{\mu\nu}(l_{\mu\nu},\phi) \right)]
\, ,
\label{MDM-Action}
\end{equation}
where $\kappa = 8 \pi G$. One can show that the action (\ref{MDM-Action})
takes the following form \cite{Chamseddine:2013kea}
\begin{equation}
S_1 [l_{\mu\nu},\phi, \Phi, \lambda] 
 = 
- \frac{1}{2 \kappa}\int d^4 x \sqrt{- l }
\left[
\Phi^2 R \left( l_{\mu\nu} \right)
+ 6 l^{\mu \nu} \nabla_{\mu } \Phi \nabla_{\nu } \Phi
- 
\lambda
\left(
\Phi^2
-
l^{\mu \nu}
\partial_{\mu} \phi \partial_{\nu} \phi
\right)
\right]
\, ,
\label{MDM-Action-1}
\end{equation}
where $l^{\mu \nu}$ denotes the inverse of the auxiliary metric
\begin{equation}
l^{\mu \nu} l_{\nu \rho} = {\delta^{\mu}}_{\rho}
\, .
\label{Auxiliary-metric-inverse}
\end{equation}
The last term from the action (\ref{MDM-Action-1}) implements the constraint that represents the consistency condition (\ref{MDM-consistency}) and $\lambda$ is the corresponding Lagrange multiplier \cite{Golovnev:2013jxa,Barvinsky:2013mea}. Although it
is possible to formulate the Einstein-Hilbert action without the constraint, the equation (\ref{MDM-Action-1}) is convenient for generalization to a FALVA model since it provides an unique procedure to extend both the action and the metric consistency constraint to the fractional fields. The mimetic gravity model has three degrees of freedom: two transverse polarizations of the gravitational field and one conformal degree of freedom of the scalar field.

In what follows, we are going to consider the general homogeneous FLRW cosmological background characterized by the following line element
\begin{equation}
d^2 s_{g} = \mathcal{N}^2 (t) dt^2 
- \mathcal{A}^2 (t) \delta_{ij} dx^{i} dx^{j}
\, .
\label{FLRW-physical}
\end{equation}
The physical metric tensor corresponding to $d^2 s_{g}$ can be obtained from non-vanishing spatially homogeneous fields $\phi (t)$ and $l_{\mu \nu} (t)$ which, at their turn, define the following auxiliary line element
\begin{equation}
d^2 s_{l} = N^2 (t) dt^2 
- a^2 (t) \delta_{ij} dx^{i} dx^{j}
\, .
\label{FLRW-auxiliary}
\end{equation}
The relation between the physical and auxiliary metrics follows from the definition (\ref{MDM-metric}) which is equivalent to the following equations
\begin{equation}
\mathcal{N}^{2} (t)  = \Phi^{2} (t) N^{2}(t)
\, ,
\qquad
\mathcal{A}^{2} (t) = \Phi^{2} (t) a^{2}(t)
\, ,
\qquad
\Phi^2 (t) = N^{-2} (t) \dot{\phi}^2 (t)
\, .
\label{FLRW-correspondence}
\end{equation}
By substituting the equations $d^2 s_{g}$ and $d^2 s_{l}$ into the action (\ref{MDM-Action-1}) the action $S_1 [l_{\mu\nu},\phi, \Phi, \lambda] $ can be put into the following form 
\begin{equation}
S_1 [N, a ,\phi, \Phi, \lambda] 
= 
- \frac{1}{2 \kappa}\int d t 
N a^3
\left[
6 \Phi^2
\left(
\frac{\ddot{a}}{N^2 a}
+
\frac{\dot{a}^2}{N^2 a^2}
- \frac{\dot{N} \dot{a}}{N^3 a}
\right)
+ 
\frac{6 \dot{\Phi}^2}{N^2}
- 
\lambda
\left(
\Phi^2
-
\frac{\dot{\phi}^2}{N^2}
\right)
\right]
\, .
\label{FLRW-action}
\end{equation}
Note that the last equation from (\ref{FLRW-correspondence}) is just 
the consistency constraint (\ref{MDM-consistency}) from above. As before, the constraint is specified by the equation of motion of the Lagrange multiplier $\lambda$. The relevant variable here is time $t$ since the spatial integration generates an overall constant volume factor that plays no role in the dynamics of fields.

It is worth mentioning that the density energy of the scalar field in the FLRW background goes with $\mathcal{A}^{-3}$. The dependence of the density energy on $\mathcal{A}$ is not affected by the presence of other matter fields. Thus, the mimetic gravity model has properties similar to the cold dark matter. For more details on the application of this model to cosmology, see e. g. \cite{Chamseddine:2014vna,Chamseddine:2016uef}.

\section{Fractional Mimetic Dark Matter Models in FALVA Formulation}

In this section, we generalize the mimetic dark matter model discussed
in the previous section to the fractional gravity. We are going to use
the FALVA formulation without fractional derivatives for which we follow \cite{Nabulsi:2008a}-\cite{Godinho:2018sxu} and we abbreviate this model by FALVA-FMDDM. Firstly, we are going to give the most general FALVA-FMDDM. Then we are going to discuss two particular models constructed in backgrounds of higher symmetry determined by particular forms of the auxiliary metric. The first particular FALVA-FMDDM has a standard non fractional mimetic field. For the second model, we consider a simple ansatz for the fractional mimetic field.

\subsection{General FALVA-FMDMM} \label{subsect-FMDMM-gen}

The basic object in the FALVA formulation of the fractional field theory is the fractional action integral. For a set of fields $\psi$ in one-dimension, the fractional action takes the following form
\begin{equation}
S^{\alpha}[\psi](t) = 
\frac{1}{\Gamma(\alpha)}
\int^{t}_{a} d\tau
L(\psi(\tau),\dot{\psi}(\tau),\ddot{\psi}(\tau),\tau)
(t-\tau)^{\alpha-1} 
\, ,
\label{FALVA-gen-int}
\end{equation}
where $\psi:[a,b] \in \mathbb{R} \rightarrow M$ satisfies the boundary conditions
$\psi(a) = \psi_{1}$ and $\psi(b) = \psi_{2}$, $\alpha$ is the fractionality parameter $\alpha \in (0,1]$ and the dot denotes
the standard derivative with respect to $\tau$ variable. Here, 
$\Gamma (\alpha) = \int_{0}^{\infty} t^{\alpha - 1} \exp (- t) \, dt$
is the Euler gamma function. Alternatively, the action $S^{\alpha}(t)$ 
can be written as follows 
\begin{equation}
S^{\alpha}[\psi](t)
=
\int_{a}^{t}
L(\psi(\tau),\dot{\psi}(\tau),\ddot{\psi}(\tau),\tau)
d \mu_{t}
\, ,
\label{FALVA-gen-int-2}
\end{equation}
where $\mu_{t}(\tau)=\frac{{1}}{{\Gamma(\alpha+1)}}\{t^{\alpha}-(t-\tau)^{\alpha}\}$ 
has a scale property that generates a memory effect. Typically, the complexity of physical systems is connected to the memory of long range interactions by non-local (spatial) effects which are incorporated in the FALVA formalism through the integration measure $d \mu_t$ \cite{Pod:2003}.

We define the FALVA-FMDMM by generalizing $S_1 [N, a ,\phi, \Phi, \lambda]$ from the equation (\ref{FLRW-action}) to the following fractional action integral
\begin{align}
S^{\alpha} [N, a ,\phi, \Phi, \lambda] (t)
& = 
- \frac{1}{2 \kappa \Gamma(\alpha)} \int^{t}_{a} d \tau 
N a^3
\left[
6 \Phi^2
\left(
\frac{\ddot{a}}{N^2 a}
+
\frac{\dot{a}^2}{N^2 a^2}
- \frac{\dot{N} \dot{a}}{N^3 a}
\right)
+
\frac{6 \dot{\Phi}^2}{N^2}
\right.
\nonumber
\\
& - 
\left.
\lambda
\left(
\Phi^2
-
\frac{\dot{\phi}^2}{N^2}
\right)
\right]
(t - \tau)^{\alpha-1}
\, .
\label{FALVA-FLRW-action}
\end{align}
The last term from $S^{\alpha} [N, a ,\phi, \Phi, \lambda]$ implements the metric constraint in the fractional model. The derivatives of the fields $\psi = \left\{ a, N, \Phi , \lambda , \phi \right\}$ are defined by the standard calculus as required by the FALVA construction. One can also interpret the equation (\ref{FALVA-FLRW-action}) as describing a set of models parametrized by the fractionality parameter $\alpha$.

The dynamics of the FALVA-FMDMM is described by the fractional variational principle applied to $S^{\alpha} [N, a ,\phi, \Phi, \lambda]$ with the variation taken over the space of the functions $\psi$ with fixed end points. Then the variational principle is equivalent to the 
fractional Euler-Lagrange equations that take the following form \cite{Godinho:2018sxu}
\begin{align}
\mathcal{D} \psi 
& = \frac{\partial \mathcal{L}}{\partial \psi} 
- \frac{d}{d \tau} \left( \frac{\partial \mathcal{L}}{\partial \dot{\psi}} \right)
+ \frac{d^2}{d \tau^2} \left( \frac{\partial \mathcal{L}}{\partial \ddot{\psi}} \right)
\nonumber
\\
& + \left( \frac{1 - \alpha}{t - \tau} \right) 
\frac{\partial \mathcal{L}}{\partial \dot{\psi}}
- 2 \left( \frac{1 - \alpha}{t - \tau} \right) 
\frac{d}{d \tau} 
\left( \frac{\partial \mathcal{L}}{\partial \ddot{\psi}} 
\right)
+ \frac{(1- \alpha)(2- \alpha )}{(t - \tau )^2} 
\frac{\partial \mathcal{L}}{\partial \ddot{\psi}}  = 0
\, ,
\label{FALVA-Eq-Motion-General}
\end{align}
By applying the equation (\ref{FALVA-Eq-Motion-General}) to the fields of the model, we obtain the following equations  
\begin{align}
\mathfrak{D} a 
& = 
3 \Phi^2 N^2 [2aN(\dot a\dot N-\ddot {a})-N\dot {a}^2]
+
\dot N-\dot \Phi N - \Phi \dot N
-
6aN^3 \Phi(\ddot a \Phi+a \ddot \Phi)
\nonumber
\\ 
& + 6 a^2 \Phi^2 \dot N N 
+ 
6 N^3 \Phi^2(\dot {a}^2+\ddot {a}a-2 \dot a) 
+
3 
\left(\dfrac{1-\alpha}{t- \tau} \right)
N^3 \Phi 
\left[
4a (\dot a \Phi+a \dot \Phi) - 2  \Phi \dot a
\right] 
\nonumber \\
&-\dfrac{3 (1-\alpha)(2-\alpha)}{(t-\tau)^2}
a^2 \Phi^2 N^3 
= 0
\, ,
\label{FALVA-eq-a}
\\
\mathfrak{D} N 
& = 
\Phi(\ddot a a +\dot {a}^2 a)
-2\dot \Phi N \dot a a^2-a N\Phi(\ddot a a +2\dot {a}^2)
-12\Phi^3-6 \lambda \Phi \phi^2
\nonumber
\\
& +
\left(\dfrac{1-\alpha}{t- \tau} \right)\dot{a} a^2
\Phi N 
= 0
\, ,  
\label{FALVA-eq-N}
\\
\mathfrak{D} \Phi 
& = 
\Phi[a(\dot N\dot a  +N \ddot a)  - N \dot {a}^2]
+
a N [3\dot a \dot \Phi+a \ddot \Phi-\dot N a \Phi]
-
\left(\dfrac{1-\alpha}{t- \tau} \right) 
a^2 \Phi N^3
= 0\, ,  
\label{FALVA-eq-Phi}
\\
\mathfrak{D} \lambda 
& =
 \Phi^2
-
\frac{\dot{\phi}^2}{N^2} = 0
\, ,
\label{FALVA-eq-lambda}
\\
\mathfrak{D} \phi 
& =
6 \lambda \dot{a} \dot{\phi} \phi N
+ \lambda a \ddot{\phi} \phi N
+ \lambda a \dot{\phi}^2 N
+ \dot{\lambda} a \dot{\phi} \phi N
- \lambda a \dot{\phi} \phi \dot{N}
+
\left(\dfrac{1-\alpha}{t- \tau} \right) 
\lambda a \dot{\phi} \phi N 
= 0
\, .
\label{FALVA-eq-phi}
\end{align}
Note that the non-linear differential equations (\ref{FALVA-eq-a}) - (\ref{FALVA-eq-phi}) are defined in the standard calculus. These equations are non-local and they contain damping coefficients that determine the field dynamics. The equation (\ref{FALVA-eq-lambda}) is the constraint on $\Phi$ which is a consequence of the metric structure of the original 
mimetic model that is assumed to hold in the fractional case.

\subsection{FALVA-FMDMM with non-fractional mimetic field} \label{subsect-FMDMM-nf}

In general, the set of non-linear equations (\ref{FALVA-eq-a}) - (\ref{FALVA-eq-phi}) is difficult to solve. However, some simplifications can be made for models with specific properties, e. g. for higher symmetric backgrounds. As an example, consider the backgrounds with constant auxiliary Hubble parameter $\dot{a}/a = H_0$ and homogeneous field $\phi (x) = \phi (t)$. By implementing these properties at the Lagrangian level and after solving the consistency condition which fixes $ \phi (t)  = t + C$, where $C$ is a constant which we set to zero, we obtain the following Lagrangian
\begin{equation}
\mathcal{L} = - \frac{3 \, a^3}{\kappa \Gamma (\alpha ) N^3 }  
\left[ 
H_0 \left( 
2 H_0 -  \frac{\dot{N}}{N} 
\right) 
+ \frac{\dot{N}^2}{N^2} 
\right]
\left( t - \tau \right)^{\alpha - 1}
\, . 
\label{FALVA-sol-Lag}
\end{equation}
The degrees of freedom of this model are the fields $a(\tau )$ and $N ( \tau ) $ with $\phi = \tau$. The field $a (\tau )$ does not enter the Lagrangian dynamically. However, its form is fixed by the auxiliary Hubble parameter condition to $a(\tau ) = \exp (H_0 \tau )$. Since 
$a(\tau )$ is determined by a condition which is not an equation of motion,  we can use the exponential function to fix the form of the Lagrangian further to 
\begin{equation}
\mathcal{L} = - \frac{3\,  e^{3 H_0 \tau} \left( t - \tau \right)^{\alpha - 1}}{\kappa \Gamma (\alpha ) N^3}
\left[ 
H_0 \left( 
2 H_0 -  \frac{\dot{N}}{N} 
\right) 
+ \frac{\dot{N}^2}{N^2} 
\right]
\, . 
\label{FALVA-sol-Lag-1}
\end{equation}
The equation of motion $\mathcal{D} N = 0 $ can be obtained by using the 
formula (\ref{FALVA-Eq-Motion-General}). After some calculations, we are led to the following non-linear first order differential equation
\begin{equation}
\mathcal{D} N = 2 \frac{\ddot{N}}{N} - 3 \frac{\dot{N}^2}{N^2} +
\left(
\frac{1 - \alpha}{t - \tau} + 8 H_0 - 6 
\right)
\frac{\dot{N}}{N}
-
\left(
\frac{1 - \alpha}{t - \tau} + 3 H^2_0 - 6 H_0
\right)
= 0
\, .
\label{FALVA-sol-eq-N}
\end{equation}
By making the substitution $F (\tau ; \alpha) = \dot{N}/N$, the equation  (\ref{FALVA-sol-eq-N}) is transformed into the following Riccati equation
\begin{equation}
\dot{F} (\tau ; \alpha) - \frac{1}{2} F^2 (\tau ; \alpha) + 
\left(
\frac{1 - \alpha}{t - \tau} + 8 H_0 - 6 
\right)
F (\tau ; \alpha)
-
\frac{1}{2}
\left(
\frac{1 - \alpha}{t - \tau} + 3 H^2_0 - 6 H_0
\right)
= 0
\, .
\label{FALVA-sol-eq-Riccati}
\end{equation}
The equation (\ref{FALVA-sol-eq-Riccati}) can be converted into a second-order linear ordinary differential equation by introducing the following function 
\begin{equation}
u (\tau ; \alpha) 
= \exp \left( - \frac{1}{2} \int d \tau F (\tau ; \alpha) \right)
\, .
\label{FALVA-sol-Riccati-var-linear}
\end{equation}
After substituting the equation (\ref{FALVA-sol-Riccati-var-linear}) in to the equation (\ref{FALVA-sol-eq-Riccati}) and after some algebraic manipulations we obtaine the following final equation
\begin{align}
\ddot{u} - \left( \frac{1 - \alpha}{\tau - t} - 8 H_0 + 6 \right) \dot{u}
+
\left( \frac{\alpha - 1}{\tau - t} + 3 H^2_0 - 6 H_0 \right) = 0
\, .
\label{FALVA-sol-eq-Riccati-linear}
\end{align}
The general solution of the equation (\ref{FALVA-sol-eq-Riccati-linear}) can be found by standard methods and it is has the form
\begin{align}
u(\tau ; \alpha) & = 
\int^{\tau}_{1}
d \xi \, 
\left\{
\frac{e^{2 (4 H_0 - 3)(t - \xi)}}{2^{\alpha + 2}}
\left( t - \xi \right)
\left[
2\left( \alpha - 1 \right) \left(4 H_0 - 3 \right)
\Gamma \left( \alpha - 1 , 2 (4 H_0 - 3) (t - \xi ) \right)
\right.
\right.
\nonumber
\\
& - 
\left.
\left.
3 \left( H_0 - 2 \right) H_0
\Gamma \left( \alpha , 2 (4 H_0 - 3 )(t - \xi ) \right)
\right]
\left[
\left( 4 H_0 - 3 \right) \left( t - \xi \right)
\right]^{- \alpha}
\right.
\nonumber
\\
& +
\left.
e^{2(4 H_0 - 3 )(t - \xi ) - (\alpha - 1) \ln (t - \xi )}
C_1
\right\}
+ C_2
\, .
\label{FALVA-sol-u-u}
\end{align}
Here, $\Gamma (a,x)$ is the incomplete gamma function 
\begin{equation}
\Gamma (a,x)=\int _{x}^{\infty }t^{a-1}\,e^{-t}\,dt,
\, ,
\label{FALVA-incomplete-gamma}
\end{equation}
and $C_1$ and $C_2$
are integration constants. The solution of the equation $ \mathcal{D} N = 0 $ is written as
\begin{equation}
N (\tau ; \alpha) = u^{-2} (\tau ; \alpha)
\, .
\label{FALVA-sol-N-u}
\end{equation}
The above equation shows that the function $u(\tau ; \alpha)$ determines a family of fractional auxiliary backgrounds parametrized by
the parameter $\alpha$.  

Let us discuss the function $u(\tau ; \alpha)$. To simplify the formulas, we introduce the variable $z = 2 (4 H_0 - 3)(t - \xi)$ which is positive for $ t > \xi$ and $H_0 > 3/4$ or $ t < \xi $ and $ H_0 < 3/4$. At $ H_0 = 3/4$, the change of variable $\xi \to z$ is not well defined.  The equation (\ref{FALVA-sol-u-u}) takes the following form in terms of $z$ variable
\begin{align}
u(\tau ; \alpha) & = 
-\frac{1}{2 (4 H_0 - 3)}
\int^{b}_{a}
d z \, 
\left\{
\frac{e^{z} z^{1- \alpha}}{8 (4 H_0 - 3)}
\left[
2\left( \alpha - 1 \right) \left(4 H_0 - 3 \right)
\Gamma \left( \alpha - 1 , z \right)
\right.
\right.
\nonumber
\\
& - 
\left.
\left.
3 \left( H_0 - 2 \right) H_0
\Gamma \left( \alpha , z \right)
\right]
+
C_1 e^{z}
\left[
\frac{z}{2(H_0 - 3)}
\right]^{1-\alpha}
\right\}
+ C_2
\, ,
\label{FALVA-sol-u-u-1}
\end{align}
where the integration limits are $a = 2 (4 H_0 - 3)(t-1)$ and 
$b =  2 (4H_0 - 3) (t-\tau )$ and $C_1$ and $C_2$ are real constants. 

At small values of $z$, one can approximate $\Gamma (\alpha, z ) / z^{\alpha} \sim - \frac{1}{\alpha}$ since $\alpha < 1$ \cite{Gradshteyn:2015}. Then
$u (\tau;\alpha)$ is given by the following formula at lowest order in $z$
\begin{align}
u_0 (\tau , \alpha) 
&
\simeq
\frac{e^{2 (4 H_0 - 3)(t-1)}}{8 (4 H_0 - 3)}
\left\{
\exp \left( \frac{t-\tau }{t-1} \right) - 1
- 
\frac{3(H_0 - 2)H_0}{2(4H_0 - 3) \alpha}
\left[
	\exp \left( \frac{t-\tau }{t-1} \right) 
		2 (4H_0 - 3) (t-\tau ) 
\right.
\right.
\nonumber
\\
& -
\left.
\left. 
		\exp \left( \frac{t-\tau }{t-1} \right)
		-		
		2 (4 H_0 - 3)(t-1)
		 + 1
\right]
\right\}
 + C
\, .
\label{FALVA-sol-uu-0}
\end{align}
where $C$ is a constant.

Another approximation can be made at large values of $z$ where the equation (\ref{FALVA-sol-u-u}) is approximated by the following one
\begin{align}
u_1(\tau ; \alpha) & \simeq
- \frac{1}{8 ( 4 H_0 - 3 )} \ln 
\left( 
\frac{t-\tau }{t-1} 
\right)
- 
\frac{3 (H_0 - 2) H_0}{8 ( 4 H_0 - 3 )} 
\left(
1-\tau 
\right)
\nonumber
\\
& 
- C_1
2^{\alpha - 6} (4 H_0 - 3)^{2 \alpha - 4}
\left[
e^{2(4 H_0 - 3)(t - \tau )} (t - \tau )^{\alpha - 1}
-
e^{2(4 H_0 - 3)(t - 1 )} (t - 1 )^{\alpha - 1}
\right]
+ C_2
\, ,
\label{FALVA-sol-uu-infty}
\end{align}
with $C_1$ and $C_2$ are constants. However, one should note that large
$z$ is equivalent with large $t - \xi$ and $\xi \in [1, \tau ]$ with
$\tau \in [1, t ]$. Recall that the initial action is defined at a fixed value of $t$, therefore the formula (\ref{FALVA-sol-uu-infty}) is interesting in the
limit $\lim_{t \to \infty} S^{\alpha}[\psi](t)$.

\subsection{FALVA-FMDMM with a fractional mimetic field} \label{subsect-FMDMM-f}

The FALVA-FMDMM from the previous section has a non-fractional field $\phi$ which is the same as in the standard mimetic dark model. The simplest model with fractional mimetic field can be obtained by taking the following ansatz $\phi (t) = t^{\alpha + 1}$. Since the constraint (\ref{MDM-consistency}) is the same, it follows that 
\begin{equation}
\mathcal{N} (t) =  \pm (\alpha + 1) t^{\alpha}
\, .
\label{FALVA-model-2-consistency}
\end{equation}
Thus, we can take $\Phi (t) = \pm (\alpha+1) t^{\alpha} N^{-1} (t)$ which leaves the field $N$ the only degree of freedom of the model. Its dynamics follows from the Lagrangian
\begin{equation}
\mathcal{L} = - \frac{3 \left( \alpha + 1 \right)^{2}}{\kappa \Gamma (\alpha)}
\left( 
\frac{e^{H_0 \tau}}{ N}
\right)^3
\left[
- H_0 \frac{\dot{N}}{N} + \frac{\alpha^2}{\tau^2} + 2 H^{2}_{0}
\right]
\tau^{2 \alpha}
(t - \tau )^{\alpha - 1}
\, .
\label{FALVA-model-2-Lagrangian}
\end{equation}
By applying the fractional variational principle, we obtain the following equation of motion
\begin{equation}
\frac{\dot{N}}{N} - \frac{3 \alpha^2}{H_0 \tau^2} - \frac{2 \alpha}{\tau} + 6 H_0 - 3 = 0
\, .
\label{FALVA-model-2-eq-motion}
\end{equation}
This is an ordinary differential equation that has the following solution
\begin{equation}
N(\tau ; \alpha ) = C_3 \, \tau^{- 2 \alpha} \exp 
\left[ 
- \frac{3 \alpha}{H_0 \tau} + 3 (2 H_0 - 1) \tau
\right]
\, ,
\label{FALVA-model-2-solution}
\end{equation}
where $C_3$ is a real integration constant. Then the conformal factor is
\begin{equation}
\Phi(\tau ; \alpha ) = \pm C_3 \left( \alpha + 1 \right) \tau^{- \alpha} 
\exp 
\left[ 
- \frac{3 \alpha}{H_0 \tau} + 3 (2 H_0 - 1) \tau 
\right]
\, .
\label{FALVA-model-2-conformal-factor}
\end{equation}

The solutions of the equations of motion obtained above allow one to explore the physical properties of these models as well as the consequences of the fractionality on the properties of space-time, which we are going to do in the next section.

\section{Geometrical Properties of FALVA-FMDMMs}

As we have seen in the previous section, the fractional character of the FMDMMs is defined primarily by the fractional gravity in the FALVA construction. As such, we expect that the geometry of the space-time be fractional in the sense that it should depend on the fractionality parameter.
Another source of fractionality of geometry is the mimetic field which determines the physical metric. Above, we have seen that at least for two simplified models with non-fractional and fractional mimetic fields, respectively, the dynamics of the
degrees of freedom can be completely determined. This is possible since the equations of motion become tractable at the cost of additional assumptions on the fields. 

In this section, we are going to
investigate some geometrical and physical properties of these two models.
We focus mainly on the fractional aspect of the geometrical tensors which
are defined by the standard differential calculus. These definitions need not be modified since the FALVA construction is based on the standard calculus. Therefore, we can say that, in a general sense, the geometries
of each model belong to new classes of geometries, the novelty being the parametrization by the fractionality parameter.

\subsection{Geometry of FALVA-FMDMM with non-fractional mimetic field}

The mimetic field of the model given in subsection \ref{subsect-FMDMM-nf}
is the same as in the mimetic dark matter extension of the General Relativity. 
By substituting the equation (\ref{FALVA-sol-N-u}) into the metric, we conclude that the mimetic field $\phi$ induces the following space-time metric
\begin{equation}
ds^{2}_{g} = - dt^2 + e^{2 H_0 t} u^4 (\tau ; \alpha) \delta_{ij} dx^i dx^j
\, ,
\label{FALVA-FMDMM-1-metric}
\end{equation} 
where the general form of $u(\tau ; \alpha)$ is given by the equation
(\ref{FALVA-sol-u-u-1}). From the equation (\ref{FALVA-FMDMM-1-metric}), 
we can derive the main geometrical objects of the family of fractional space-time geometries parametrized by $\alpha$. In order to simplify the notation,
we drop the time and fractionality parameter from formulas.  

The non-vanishing Christoffel symbols have the following form
\begin{align}
{\Gamma^{0}}_{1 1} & = 
 e^{2 H_0 t} u^3 \left(H_0 u+2 \dot{u}\right)
\, ,
\label{FALVA-FMDMM-1-chris-1}
\\
 {\Gamma^{0}}_{2 2} & = 
 e^{2 H_0 t} u^3 \left(H_0 u+2 \dot{u}\right)
 \, ,
\label{FALVA-FMDMM-1-chris-2} 
\\
 {\Gamma^{0}}_{3 3} & = 
 e^{2 H_0 t} u^3 \left(H_0 u+2 \dot{u}\right)
 \, ,
\label{FALVA-FMDMM-1-chris-3} 
\\
{\Gamma^{1}}_{1 0} & = 
 H_0 +\frac{2 \dot{u}}{u}
 \, ,
\label{FALVA-FMDMM-1-chris-4} 
\\
 {\Gamma^{2}}_{2 0} & = 
 H_0 +\frac{2 \dot{u}}{u}
 \, ,
\label{FALVA-FMDMM-1-chris-5} 
\\
 {\Gamma^{3}}_{3 0} & = 
 H_0 +\frac{2 \dot{u}}{u}
 \, .
\label{FALVA-FMDMM-1-chris-6} 
\end{align}
By using the Christoffel symbols, we can calculated the components of 
the Riemann tensor and we obtain the following results
\begin{align}
{R^0}_{1 1 0} & = 
-e^{2 H_{0} t} u^2 \left(H^{2}_{0} u^2+2 \dot{u}^2+2 u \left(2 H_{0} \dot{u}+\ddot{u}\right)\right) 
\, ,
\label{FALVA-FMDMM-1-riem-1} 
\\
{R^0}_{2 2 0} & = -e^{2 H_{0} t} u^2 \left(H^{2}_{0} u^2+2 \dot{u}^2+2 u \left(2 H_{0} \dot{u}+\ddot{u}\right)\right) 
\, ,
\label{FALVA-FMDMM-1-riem-2} 
\\
{R^0}_{3 3 0} & = -e^{2 H_{0} t} u^2 \left(H^{2}_{0} u^2+2 \dot{u}^2+2 u \left(2 H_{0} \dot{u}+\ddot{u}\right)\right) 
\, ,
\label{FALVA-FMDMM-1-riem-3} 
\\
{R^1}_{0 1 0} & = -H^{2}_{0}-\frac{2 \dot{u}^2}{u^2}-\frac{2 \left(2 H_{0} \dot{u}+\ddot{u}\right)}{u} 
\, ,
\label{FALVA-FMDMM-1-riem-4} 
\\
{R^1}_{2 2 1} & = -e^{2 H_{0} t} u^2 \left(H_{0} u+2 \dot{u}\right)^2 
\, ,
\label{FALVA-FMDMM-1-riem-5} 
\\
{R^1}_{3 3 1} & = -e^{2 H_{0} t} u^2 \left(H_{0} u+2 \dot{u}\right)^2 
\, ,
\label{FALVA-FMDMM-1-riem-6} 
\\
{R^2}_{0 2 0} & = -H^{2}_{0}-\frac{2 \dot{u}^2}{u^2}-\frac{2 \left(2 H_{0} \dot{u}+\ddot{u}\right)}{u} 
\, ,
\label{FALVA-FMDMM-1-riem-7} 
\\
{R^2}_{1 2 1} & = e^{2 H_{0} t} u^2 \left(H_{0} u+2 \dot{u}\right)^2 
\, ,
\label{FALVA-FMDMM-1-riem-8} 
\\
{R^2}_{3 3 2} & = -e^{2 H_{0} t} u^2 \left(H_{0} u+2 \dot{u}\right)^2 
\, ,
\label{FALVA-FMDMM-1-riem-9} 
\\
{R^3}_{0 3 0} & = -H^{2}_{0}-\frac{2 \dot{u}^2}{u^2}-\frac{2 \left(2 H_{0} \dot{u}+\ddot{u}\right)}{u} 
\, ,
\label{FALVA-FMDMM-1-riem-10} 
\\
{R^3}_{1 3 1} & = e^{2 H_{0} t} u^2 \left(H_{0} u+2 \dot{u}\right)^2 
\, ,
\label{FALVA-FMDMM-1-riem-11} 
\\
{R^3}_{2 3 2} & = e^{2 H_{0} t} u^2 \left(H_{0} u+2 \dot{u}\right)^2 
\, .
\label{FALVA-FMDMM-1-riem-12} 
\end{align}
The next tensor of interest is the Ricci tensor. Its non-vanishing components are
\begin{align}
R_{0 0} & = 
-3 H^{2}_{0}-\frac{6 \dot{u}^2}{u^2}-\frac{6 \left(2 H_{0} \dot{u}+\ddot{u}\right)}{u}
\, , 
\label{FALVA-FMDMM-1-ricci-1} 
\\
R_{1 1} & = e^{2 H_{0} t} u^2 \left(3 H^{2}_{0} u^2+10 \dot{u}^2+2 u \left(6 H_{0} \dot{u}+\ddot{u}\right)\right) 
\, , 
\label{FALVA-FMDMM-1-ricci-2} 
\\
R_{2 2} & = e^{2 H_{0} t} u^2 \left(3 H^{2}_{0} u^2+10 \dot{u}^2+2 u \left(6 H_{0} \dot{u}+\ddot{u}\right)\right) 
\, , 
\label{FALVA-FMDMM-1-ricci-3} 
\\
R_{0 0} & = e^{2 H_{0} t} u^2 \left(3 H^{2}_{0} u^2+10 \dot{u}^2+2 u \left(6 H_{0} \dot{u}+\ddot{u}\right)\right)
\, . 
\label{FALVA-FMDMM-1-ricci-4} 
\end{align}
It is easy to see that the Ricci scalar has the following form
\begin{equation}
R =
\frac{12 
\left[
H^{2}_{0} u^2+3 \dot{u}^2 + u 
\left(4 H_0 \dot{u} + \ddot{u} 
\right)
\right]}{u^2}
\, .
\label{FALVA-FMDMM-1-ricci-scalar} 
\end{equation}
The geodesic equations can be calculated from the Christoffels symbols. To this end, we introduce the components of the velocity $v^{\mu}(s)= dx^{\mu}(s)/ds$
along the geodesics as functions on the proper time $s$. Then we get the following set of equations
\begin{align}
\frac{d v^{0}}{d s} & = 
- e^{2 H_{0} t} 
\left[(v^1)^2+(v^2)^2+(v^3)^2\right] 
u^4 \left(H_{0}+\frac{2 \dot{u}}{u} \right)
\, ,
\label{FALVA-FMDMM-1-geo-1}  
\\
\frac{d v^{1}}{d s} & =
-2 v^0 v^1 \left(H_{0}+\frac{2 \dot{u}}{u}\right)
\, ,
\label{FALVA-FMDMM-1-geo-2} 
\\
\frac{d v^{2}}{d s} & =
-2 v^0 v^2 \left(H_{0}+\frac{2 \dot{u}}{u}\right) 
\, ,
\label{FALVA-FMDMM-1-geo-3}
\\
\frac{d v^{3}}{d s} & =
-2 v^0 v^3 \left(H_{0}+\frac{2 \dot{u}}{u}\right)
\, .
\label{FALVA-FMDMM-1-geo-4}
\end{align}

The above relations have been obtained for the general $u(\tau; \alpha)$. When the explicit time-dependence of $u$ on $\tau$ is plugged-in, concrete, but not very illuminating formulas, are obtained. 

\subsection{Geometry of FALVA-FMDMM with fractional mimetic field}

The FALVA-FMDMM with fractional mimetic field given by the Lagrangian (\ref{FALVA-model-2-Lagrangian}) induces the following fractional metric in the physical space-time
\begin{equation}
ds^{2}_{g}  = 
(1+\alpha )^2  t^{2 \alpha } 
\left[
- dt^2
+
C_{3}^{-2} 
t^{4 \alpha } 
e^{ 
(6 - 10 H_0 ) t + \frac{6 \alpha }{H_0 t}
}
\, 
\delta_{ij} d x^{i} dx^{j}
\right]
\, .
\label{FALVA-FMDMM-2-metric}
\end{equation}
Here, we have express the results in terms of the physical time. For simplicity of notation, we omit the dependence on time and fractionality parameter in the components of the computed quantities and we use the Cartesian coordinates. 

As in the previous case, the first geometrical object to be computed are the Christoffel symbols ${\Gamma^{\mu}}_{\nu \rho}$ corresponding to the metric (\ref{FALVA-FMDMM-2-metric}). The non-vanishing components of ${\Gamma^{\mu}}_{\nu \rho}$ are
\begin{align}
{\Gamma^{0}}_{00 } & = \frac{\alpha }{t} 
\, ,
\label{FALVA-FMDMM-2-Chris-1}
\\
{\Gamma^{0}}_{1 1} & = \frac{3 e^{6 \left(t-2 H_0  t+\frac{\alpha }{H_0 t}\right)} e^{2 H_0 t} t^{-1+4 \alpha } (\alpha +(t-2 H_0  t) )}{C_{3}^{2}}
\, ,
\label{FALVA-FMDMM-2-Chris-2}
\\
{\Gamma^{0}}_{2 2} & = \frac{3 e^{6 \left(t-2 H_0  t+\frac{\alpha }{H_0 t}\right)} e^{2 H_0 t} t^{-1+4 \alpha } (\alpha +(t-2 H_0  t) )}{C_{3}^{2}}
\, ,
\label{FALVA-FMDMM-2-Chris-3}
\\
{\Gamma^{0}}_{3 3} & = \frac{3 e^{6 \left(t-2 H_0  t+\frac{\alpha }{H_0 t }\right)} e^{2 H_0 t} t^{-1+4 \alpha } (\alpha +(t-2 H_0  t) )}{C_{3}^{2}}
\, ,
\label{FALVA-FMDMM-2-Chris-4}
\\
{\Gamma^{1}}_{1 0} & = \frac{3 (\alpha +(t-2 H_0  t) )}{t} 
\, ,
\label{FALVA-FMDMM-2-Chris-5}
\\
{\Gamma^{2}}_{2 0} & = \frac{3 (\alpha +(t-2 H_0  t) )}{t} 
\, ,
\label{FALVA-FMDMM-2-Chris-6}
\\
{\Gamma^{3}}_{3 0} & = \frac{3 (\alpha +(t-2 H_0  t))}{t} 
\, ,
\end{align}
where, as usual, the Christoffel symbols are symmetric in the lower indices. It follows that the corresponding Riemann tensor has the following non-vanishing components
\begin{align}
{R^0}_{1 1 0} & =  
-\frac{3 e^{6 \left(t-2 H_0  t+\frac{\alpha }{H_0 t}\right)} e^{2 H_0 t} t^{-2+4 \alpha } \left(\alpha  (-1+2 \alpha
)+5 (1-2 H_0 ) t \alpha + 3 (1-2 H_0 )^2 t^2 \right)}{C^{2}_{3}} 
\, ,
\label{FALVA-FMDMM-2-Riemann-1}
\\
{R^0}_{2 2 0} & = 
-\frac{3 e^{6 \left(t-2 H_0  t+\frac{\alpha }{H_0 t}\right)} e^{2 H_0 t} t^{-2+4 \alpha } \left(\alpha  (-1+2 \alpha
)+5 (1-2 H_0 ) t \alpha + 3 (1-2 H_0 )^2 t^2 \right)}{C^{2}_{3}} 
\, ,
\label{FALVA-FMDMM-2-Riemann-2}
\\
{R^0}_{3 3 0} & = 
-\frac{3 e^{6 \left(t-2 H_0  t+\frac{\alpha }{H_0 t}\right)} e^{2 H_0 t} t^{-2+4 \alpha } \left(\alpha  (-1+2 \alpha
)+5 (1-2 H_0 ) t \alpha  + 3 (1-2 H_0 )^2 t^2 \right)}{C^{2}_{3}} 
\, ,
\label{FALVA-FMDMM-2-Riemann-3}
\\
{R^1}_{0 1 0} & = 
\frac{3 \left(\alpha -2 \alpha ^2+5 (-1+2 H_0 ) t \alpha  
- 3 (1-2 H_0 )^2 t^2 \right)}{t^2} 
\, ,
\label{FALVA-FMDMM-2-Riemann-4}
\\
{R^1}_{2 2 1} & = 
-\frac{9 e^{6 \left(t-2 H_0  t+\frac{\alpha }{H_0 t}\right)} e^{2 H_0 t} t^{-2+4 \alpha } (\alpha +(t-2 H_0  t))^2}{C^{2}_{3}}
\, ,
\label{FALVA-FMDMM-2-Riemann-5}
\\
{R^1}_{3 3 1} & = 
-\frac{9 e^{6 \left(t-2 H_0  t+\frac{\alpha }{H_0 t}\right)} e^{2 H_0 t} t^{-2+4 \alpha } (\alpha +(t-2 H_0  t))^2}{C^{2}_{3}}
\, ,
\label{FALVA-FMDMM-2-Riemann-6}
\\
{R^2}_{0 2 0} & = 
\frac{3 \left(\alpha -2 \alpha ^2+5 (-1+2 H_0 ) t \alpha 
- 3 (1-2 H_0 )^2 t^2 \right)}{t^2} 
\, ,
\label{FALVA-FMDMM-2-Riemann-7}
\\
{R^2}_{1 2 1} & = 
\frac{9 e^{6 \left(t-2 H_0  t+\frac{\alpha }{H_0 t}\right)} e^{2 H_0 t} t^{-2+4 \alpha } (\alpha +(t-2 H_0  t))^2 }{C^{2}_{3}}
\, ,
\label{FALVA-FMDMM-2-Riemann-8}
\\
{R^2}_{3 3 2} & = 
-\frac{9 e^{6 \left(t-2 H_0  t+\frac{\alpha }{H_0 t}\right)} e^{2 H_0 t} t^{-2+4 \alpha } (\alpha +(t-2 H_0  t))^2}{C^{2}_{3}}
\, ,
\label{FALVA-FMDMM-2-Riemann-9}
\\
{R^3}_{0 3 0} & = 
\frac{3 \left(\alpha -2 \alpha ^2+5 (-1+2 H_0 ) t \alpha - 
3 (1-2 H_0 )^2 t^2 \right)}{t^2} 
\, ,
\label{FALVA-FMDMM-2-Riemann-10}
\\
{R^3}_{1 3 1} & = 
\frac{9 e^{6 \left(t-2 H_0  t+\frac{\alpha }{H_0 t}\right)} e^{2 H_0 t} t^{-2+4 \alpha } (\alpha +(t-2 H_0  t) )^2}{C^{2}_{3}}
\, ,
\label{FALVA-FMDMM-2-Riemann-11}
\\
{R^3}_{2 3 2} &  = 
\frac{9 e^{6 \left(t-2 H_0  t+\frac{\alpha }{H_0 t}\right)} e^{2 H_0 t} t^{-2+4 \alpha } (\alpha +(t-2 H_0  t) )^2}{C^{2}_{3}}
\, .
\label{FALVA-FMDMM-2-Riemann-12}
\end{align}
From these, we obtain the following of the Ricci tensor
\begin{align}
R_{0 0} & = 
\frac{9 \left(\alpha -2 \alpha ^2+5 (-1+2 H_0 ) t \alpha 
- 3 (1-2 H_0 )^2 t^2 \right)}{t^2} 
\label{FALVA-FMDMM-2-Ricci-1}
\, ,
\\
R_{1 1} & = 
\frac{3 e^{6 \left(t-2 H_0  t+\frac{\alpha }{H_0 t}\right)} 
e^{2 H_0 t} t^{-2+4 \alpha } \left(\alpha  (-1+8 \alpha )+17 (1-2
H_0 ) t \alpha  +
9 (1-2 H_0 )^2 t^2 \right)}{C^{2}_{3}} 
\label{FALVA-FMDMM-2-Ricci-2}
\, ,
\\
R_{2 2} & =
\frac{3 e^{6 \left(t-2 H_0  t+\frac{\alpha }{H_0 t}\right)} e^{2 H_0 t} t^{-2+4 \alpha } \left(\alpha  (-1+8 \alpha )+17 (1-2
H_0 ) t \alpha  + 
9 (1-2 H_0 )^2 t^2 \right)}{C^{2}_{3}} 
\, ,
\label{FALVA-FMDMM-2-Ricci-3}
\\
R_{3 3} & = \frac{3 e^{6 \left(t-2 H_0  t +\frac{\alpha}{H_0 t}\right)} e^{2 H_0 t} t^{-2+4 \alpha } \left(\alpha  (-1+8 \alpha )+17 (1-2
H_0 ) t \alpha  + 
9 (1-2 H_0 )^2 t^2 \right)}{C^{2}_{3}} 
\, .
\label{FALVA-FMDMM-2-Ricci-4}
\end{align}
The non-diagonal components vanish $R_{\mu \nu} = 0$, for $\mu \neq \nu$.
By contracting with the inverse metric of (\ref{FALVA-FMDMM-2-metric}) we obtain the following Ricci scalar
\begin{equation}
R = (1+\alpha)^{-2}
\left[
18 t^{-2 (1+\alpha )} 
\left(
\alpha (-1+5 \alpha )+ 11 (1-2 H_0 ) t \alpha  + 6 (1-2 H_0 )^2 t^2 
\right)
\right]
\, .
\label{FALVA-FMDDM-2-Ricci-scalar}
\end{equation}
The geodesic equations form the following system of non-linear differential equations
\begin{align}
\frac{d u^0}{d s} & = 
\frac{e^{-12 H_0  t} 
\left[
3 e^{6 \left(t+\frac{\alpha }{H_0 t}
\right)} 
e^{2 H_0 t} (-1+2 H_0 ) t^{1+4 \alpha} 
\left(
(u^1)^2+(u^2)^2+(u^3)^2
\right)
\right.}{C^{2}_{3}\,  t}
\nonumber
\\
& -
\frac{\alpha  
\left.
\left(C
^{2}_{3} e^{12 H_0  t} (u^{0})^2 + 
3 e^{6 \left(t+\frac{\alpha }{H_0 t}\right)} 
e^{2 H_0 t} t^{4 \alpha } 
\left(
(u^1)^2+(u^2)^2+(u^3)^2
\right)\right)
\right]}{C^{2}_{3}\,  t} 
\, ,
\label{FALVA-FMDDM-2-geo-1}
\\
\frac{d u^1}{d s} & = 
-\frac{6 (\alpha +(t-2 H_0  t)) u^0 u^1}{t} 
\, ,
\label{FALVA-FMDDM-2-geo-2}
\\
\frac{d u^2}{d s} & = 
-\frac{6 (\alpha +(t-2 H_0  t) ) u^0 u^2}{t} 
\, ,
\label{FALVA-FMDDM-2-geo-3}
\\
\frac{d u^3}{d s} & = 
 -\frac{6 (\alpha +(t-2 H_0  t)) u^0 u^3}{t} 
\, .
\label{FALVA-FMDDM-2-geo-4} 
\end{align}

Note that the geometries described in this section have a fractional character manifest by the presence of the fractionality parameter $\alpha$. One can easily see that at extreme fractionality $\alpha=0$ the Christoffel symbols take the following form
\begin{align}
{\Gamma^{0}}_{00 } & = 0 
\, ,
\label{FALVA-FMDMM-2-Chris-1-alpha-0}
\\
{\Gamma^{0}}_{1 1} & = {\Gamma^{0}}_{2 2} = {\Gamma^{0}}_{3 3} 
=
\frac{3 e^{t\left( 6  - 10 H_0  \right)} (1-2 H_0)}{C_{3}^{2}}
\, ,
\label{FALVA-FMDMM-2-Chris-2-alpha-0}
\\
{\Gamma^{1}}_{1 0} & = {\Gamma^{2}}_{2 0} = {\Gamma^{3}}_{3 0} = 3 - 6 H_0 
\label{FALVA-FMDMM-2-Chris-3-alpha-0}
\, .
\end{align} 
For $\alpha = 0$, the Riemann tensor takes the simpler form
\begin{align}
{R^0}_{1 1 0} & =  {R^0}_{2 2 0} = {R^0}_{3 3 0} =
-\frac{ 9 (1-2 H_0 )^2 e^{t \left( 6 - 10 H_0 \right)}}{C^{2}_{3}} 
\, ,
\label{FALVA-FMDMM-2-Riemann-1-alpha-0}
\\
{R^1}_{0 1 0} & = {R^2}_{0 2 0} = {R^3}_{0 3 0} =
- 9 (1-2 H_0 )^2  
\, ,
\label{FALVA-FMDMM-2-Riemann-4-alpha-0}
\\
{R^1}_{2 2 1} & = {R^1}_{3 3 1} = {R^2}_{3 3 2} =
-\frac{9 e^{t \left( 6- 10 H_0 \right)}  (1-2 H_0)^2}{C^{2}_{3}}
\, ,
\label{FALVA-FMDMM-2-Riemann-5-alpha-0}
\\
{R^2}_{1 2 1} & = {R^3}_{1 3 1} = {R^3}_{2 3 2} = 
\frac{9 e^{t \left( 6- 10 H_0 \right)}  (1-2 H_0)^2}{C^{2}_{3}}
\label{FALVA-FMDMM-2-Riemann-6-alpha-0}
\, .
\end{align}
It follows that the Ricci tensor at extreme fractionality has the form
\begin{equation}
R_{0 0} = 
- 27 (1-2 H_0 )^2 
\, ,
\qquad
R_{1 1} = R_{2 2} = R_{3 3} = 
\frac{
27 e^{t( 6 - 10 H_0 )} (1-2H_0)^2}{C^{2}_{3}}
\label{FALVA-FMDMM-2-Ricci-alpha-0}
\, .
\end{equation}
Thus, we can see from the above relations, that the components of the geometric tensors are either constant or depend exponentially on time at $\alpha = 0$. At different fractionalities, the geometric tensors are given by the previous formulas.

\section{Discussions}

In this paper, we have propose a new model of non-local mimetic dark matter based on the FALVA construction. The FALVA-FMDDM generalizes previous FALVA models of gravity by including a dark matter sector. These models are in generally characterized by fractional space-time geometry. In order to understand better the space-time fractionality, we have discussed in detail two
particular cases of the general FALVA-FMDDM, with a non-fractional and a power law fractional mimetic field, respectively. In this cases, we have computed the relevant geometrical objects as given by the standard differential geometry. The results show that the fractional space-time is not a trivial generalization of the standard space-time.

While the scope of the present paper is to propose and discuss the basic geometric properties of the FALVA-FMDDM, further studies of this model are necessary in order to understand the effect of fractionality on the physical properties of the dark matter field. The most interesting problems at this moment are a deeper analysis of the fractional geometries, including the stability of the fractional space-time and the solutions of the geodesic equations and the study of the fractional mimetic dark matter in different cosmological scenarios. We hope to report on these topics in the future.   
 
\section*{Acknowledgements}

We acknowledge R. A. El-Nabulsi for very useful comments and discussions.


\begin{thebibliography}{999}



\bibitem{Nabulsi:2008a}
R.~A.~El-Nabulsi and D.F.~M.~Torres,
``Fractional action-like variational problems,''
J. Math. Phys. \textbf{49} 053521 (2008). 

\bibitem{Nabulsi:2008b}
R.~A.~El-Nabulsi, 
``Fractional field theories from multi-dimensional fractional variational problems,''
Int. J. Geom. Methods Mod. Phys. \textbf{5}, 863 (2008).

\bibitem{Nabulsi:2009c}
R.~A.~El-Nabulsi,
``The fractional calculus of variations from extended Erd\'{e}lyi-Kober
operator,''
Int. J. of Mod. Phys. B,
\textbf{16}, 3349-3361 (2009).

\bibitem{Godinho:2011}
E.~M.~C.~Abreu and C.~F.~L.~Godinho,
``Fractional Dirac Bracket and Quantization for Constrained Systems,''
Phys. Rev. E \textbf{84}, 026608 (2011).

\bibitem{Godinho:2018sxu}
C.~F.~de L.~Godinho, N.~Panza, J.~Weberszpil and J.~A.~Helay\"el-Neto,
``Variational procedure for higher-derivative mechanical models in a fractional integral,''
Euro. Phys. Lett. \textbf{129}, no.6, 60001 (2020).

\bibitem{Nabulsi:2009b}
R.~A.~El-Nabulsi,
``Fractional Dirac operators and deformed field theory on Clifford algebra,''
Chaos, Solitons and Fractals \textbf{42}, 2614 (2009).



\bibitem{El-Nabulsi:2009bup}
A.~R.~El-Nabulsi,
``Fractional Lagrangian formulation of general relativity and emergence of complex, spinorial and noncommutative gravity,''
Int. J. Geom. Meth. Mod. Phys. \textbf{6}, 25-76 (2009).

\bibitem{Vacaru:2010wn}
S.~I.~Vacaru,
``Fractional Dynamics from Einstein Gravity, General Solutions, and Black Holes,''
Int. J. Theor. Phys. \textbf{51}, 1338-1359 (2012).

\bibitem{Baleanu:2010vj}
D.~Baleanu and S.~I.~Vacaru,
``Fractional Curve Flows and Solitonic Hierarchies in Gravity and Geometric Mechanics,''
J. Math. Phys. \textbf{52}, 053514 (2011).

\bibitem{Calcagni:2011kn}
G.~Calcagni,
``Geometry of fractional spaces,''
Adv. Theor. Math. Phys. \textbf{16}, no.2, 549-644 (2012).

\bibitem{El-Nabulsi:2013mma}
R.~A.~El-Nabulsi,
``Nonstandard fractional exponential Lagrangians, fractional geodesic equation, complex general relativity, and discrete gravity,''
Can. J. Phys. \textbf{91}, no.8, 618-622 (2013).

\bibitem{El-Nabulsi:2013hsa}
A.~R.~El-Nabulsi,
``Fractional derivatives generalization of Einstein`s field equations,''
Indian J. Phys. \textbf{87}, 195-200 (2013).

\bibitem{Calcagni:2013yqa}
G.~Calcagni,
``Multi-scale gravity and cosmology,''
JCAP \textbf{12}, 041 (2013).

\bibitem{Calcagni:2021aap}
G.~Calcagni,
``Classical and quantum gravity with fractional operators,''
Class. Quant. Grav. \textbf{38}, no.16, 165005 (2021)
[erratum: Class. Quant. Grav. \textbf{38}, no.16, 169601 (2021)].

\bibitem{El-Nabulsi:2021uwp}
R.~A.~El-Nabulsi and A.~K.~Golmankhaneh,
``On fractional and fractal Einstein\textquoteright{}s field equations,''
Mod. Phys. Lett. A \textbf{36}, no.05, 2150030 (2021).


\bibitem{El-Nabulsi:2007wgc}
R.~A.~El-Nabulsi,
``Cosmology with a fractional action principle,''
Rom. Rep. Phys. \textbf{59}, 763-771 (2007).

\bibitem{El-Nabulsi:2010upv}
R.~A.~El-Nabulsi,
``Fractional action-like variational approach, perturbed Einstein's gravity and new cosmology,''
Fizika B \textbf{19}, 103-112 (2010).

\bibitem{Shchigolev:2010vh}
V.~K.~Shchigolev,
``Cosmological Models with Fractional Derivatives and Fractional Action Functional,''
Commun. Theor. Phys. \textbf{56}, 389-396 (2011).

\bibitem{Jamil:2011uj}
M.~Jamil, D.~Momeni and M.~A.~Rashid,
``Fractional Action Cosmology with Power Law Weight Function,''
J. Phys. Conf. Ser. \textbf{354}, 012008 (2012).

\bibitem{Shchigolev:2012rp}
V.~K.~Shchigolev,
``Cosmic Evolution in Fractional Action Cosmology,''
Discontinuity, Linearity and Complexity \textbf{2}, no.2, 115-123 (2013).


\bibitem{Moniz:2020emn}
P.~V.~Moniz and S.~Jalalzadeh,
``From Fractional Quantum Mechanics to Quantum Cosmology: An Overture,''
Mathematics \textbf{8}, no.3, 313 (2020).

\bibitem{Rasouli:2021lgy}
S.~M.~M.~Rasouli, S.~Jalalzadeh and P.~V.~Moniz,
``Broadening quantum cosmology with a fractional whirl,''
Mod. Phys. Lett. A \textbf{36}, no.14, 2140005 (2021).

\bibitem{Jalalzadeh:2021gtq}
S.~Jalalzadeh, F.~R.~da Silva and P.~V.~Moniz,
``Prospecting black hole thermodynamics with fractional quantum mechanics,''
Eur. Phys. J. C \textbf{81}, no.7, 632 (2021).

\bibitem{Jalalzadeh:2022uhl}
S.~Jalalzadeh, E.~W.~O.~Costa and P.~V.~Moniz,
``de Sitter fractional quantum cosmology,''
Phys. Rev. D \textbf{105}, no.12, L121901 (2022).


\bibitem{Calcagni:2021mmj}
G.~Calcagni and G.~U.~Varieschi,
``Gravitational potential and galaxy rotation curves in multi-fractional spacetimes,''
JHEP \textbf{08}, 024 (2022).

\bibitem{Rami:2015kha}
R.~A.~El-Nabulsi,
``Fractional action oscillating phantom cosmology with conformal coupling,''
Eur. Phys. J. Plus \textbf{130}, no.6, 102 (2015).

\bibitem{Shchigolev:2015}
V.~K.~Shchigolev,
``Fractional action cosmology with an effective $\Lambda$-term,''
Int. J. Adv. Astronomy  \textbf{4} 5-10 (2016).

\bibitem{Barrientos:2020kfp}
E.~Barrientos, S.~Mendoza and P.~Padilla,
``Extending Friedmann equations using fractional derivatives using a Last Step Modification technique: the case of a matter dominated accelerated expanding Universe,''
Symmetry \textbf{13}, no.2, 174 (2021).

\bibitem{Shchigolev:2021lbm}
V.~K.~Shchigolev,
``Fractional-order derivatives in cosmological models of accelerated expansion,''
Mod. Phys. Lett. A \textbf{36}, no.14, 2130014 (2021).

\bibitem{El-Nabulsi:2013mwa}
A.~R.~El-Nabulsi,
``Non-minimal coupling in fractional action cosmology,''
Indian J. Phys. \textbf{87}, 835-840 (2013).

\bibitem{El-Nabulsi:2017vmp}
R.~A.~El-Nabulsi,
``Fractional Action Cosmology with Variable Order Parameter,''
Int. J. Theor. Phys. \textbf{56}, no.4, 1159-1182 (2017).

\bibitem{El-Nabulsi:2017jss}
R.~A.~El-Nabulsi,
``Wormholes in fractional action cosmology,''
Can. J. Phys. \textbf{95}, no.6, 605-609 (2017).

\bibitem{Shchigolev:2015rei}
V.~K.~Shchigolev,
``Testing Fractional Action Cosmology,''
Eur. Phys. J. Plus \textbf{131}, no.8, 256 (2016).


\bibitem{Giusti:2020rul}
A.~Giusti,
``MOND-like Fractional Laplacian Theory,''
Phys. Rev. D \textbf{101}, no.12, 124029 (2020).

\bibitem{Varieschi:2020ioh}
G.~U.~Varieschi,
``Newtonian Fractional-Dimension Gravity and MOND,''
Found. Phys. \textbf{50}, no.11, 1608-1644 (2020)
[erratum: Found. Phys. \textbf{51}, no.2, 41 (2021)].

\bibitem{Varieschi:2020dnd}
G.~U.~Varieschi,
``Newtonian Fractional-Dimension Gravity and Disk Galaxies,''
Eur. Phys. J. Plus \textbf{136}, no.2, 183 (2021).

\bibitem{Varieschi:2020hvp}
G.~U.~Varieschi,
``Newtonian Fractional-Dimension Gravity and Rotationally Supported Galaxies,''
Mon. Not. Roy. Astron. Soc. \textbf{503}, no.2, 1915-1931 (2021).


\bibitem{Debnath:2011cj}
U.~Debnath, S.~Chattopadhyay and M.~Jamil,
``Fractional Action Cosmology: Some Dark Energy Models in Emergent, Logamediate and Intermediate Scenarios of the Universe,''
[arXiv:1107.0541 [physics.gen-ph]].

\bibitem{Pasqua:2013rza}
A.~Pasqua and S.~Chattopadhyay,
``Theoretical constraints on kinetic k-essence dark energy model in El-Nabulsi fractional action cosmology,''
Can. J. Phys. \textbf{91}, 844-849 (2013).

\bibitem{Pasqua:2013}
A.~Pasqua and S.~Chattopadhyay,
``Reconstruction of modified holographic Ricci dark energy in El-Nabulsi fractional action cosmology,''
J. Theor. App. Phys. \textbf{7} 22 (2013).

\bibitem{Pasqua:2013vfu}
A.~Pasqua and S.~Chattopadhyay,
``On the Dynamics of Dark Energy with Higher Time Derivatives of Hubble Parameter in El-Nabulsi Fractional Action Cosmology,''
Int. J. Theor. Phys. \textbf{53}, no.2, 435-448 (2014).

\bibitem{Landim:2021www}
R.~G.~Landim,
``Fractional dark energy,''
Phys. Rev. D \textbf{103}, no.8, 083511 (2021).


\bibitem{El-Nabulsi:2012wpc}
R.~A.~El-Nabulsi,
``Gravitons in fractional action cosmology,''
Int. J. Theor. Phys. \textbf{51}, 3978-3992 (2012).



\bibitem{Mashhoon:2006mj}
B.~Mashhoon,
``Toward a Nonlocal Theory of Gravitation,''
Annalen Phys. \textbf{16} 57-66 (2007).

\bibitem{Chicone:2015sda}
C.~Chicone and B.~Mashhoon,
``Nonlocal Newtonian Cosmology,''
J. Math. Phys. \textbf{57} 072501 (2016).



\bibitem{Mashhoon:2017qyw}
B.~Mashhoon,
``Nonlocal Gravity,'' Oxford University Press (2017).


\bibitem{Chamseddine:2013kea}
A.~H.~Chamseddine and V.~Mukhanov,
``Mimetic Dark Matter,''
JHEP \textbf{11}, 135 (2013).

\bibitem{Golovnev:2013jxa}
A.~Golovnev,
``On the recently proposed Mimetic Dark Matter,''
Phys. Lett. B \textbf{728}, 39-40 (2014).

\bibitem{Barvinsky:2013mea}
A.~O.~Barvinsky,
``Dark matter as a ghost free conformal extension of Einstein theory,''
JCAP \textbf{01}, 014 (2014).


\bibitem{Chaichian:2014qba}
M.~Chaichian, J.~Kluson, M.~Oksanen and A.~Tureanu,
``Mimetic dark matter, ghost instability and a mimetic tensor-vector-scalar gravity,''
JHEP \textbf{12}, 102 (2014).

\bibitem{Malaeb:2014vua}
O.~Malaeb,
``Hamiltonian Formulation of Mimetic Gravity,''
Phys. Rev. D \textbf{91}, no.10, 103526 (2015).


\bibitem{Chamseddine:2014vna}
A.~H.~Chamseddine, V.~Mukhanov and A.~Vikman,
`Cosmology with Mimetic Matter,''
JCAP \textbf{06}, 017 (2014).

\bibitem{Chamseddine:2016uef}
A.~H.~Chamseddine and V.~Mukhanov,
``Resolving Cosmological Singularities,''
JCAP \textbf{03}, 009 (2017).


\bibitem{Cognola:2016gjy}
G.~Cognola, R.~Myrzakulov, L.~Sebastiani, S.~Vagnozzi and S.~Zerbini,
``Covariant Ho\v{r}ava-like and mimetic Horndeski gravity: cosmological solutions and perturbations,''
Class. Quant. Grav. \textbf{33}, no.22, 225014 (2016).

\bibitem{Sebastiani:2016ras}
L.~Sebastiani, S.~Vagnozzi and R.~Myrzakulov,
``Mimetic gravity: a review of recent developments and applications to cosmology and astrophysics,''
Adv. High Energy Phys. \textbf{2017}, 3156915 (2017).

\bibitem{Dutta:2017fjw}
J.~Dutta, W.~Khyllep, E.~N.~Saridakis, N.~Tamanini and S.~Vagnozzi,
JCAP \textbf{02}, 041 (2018).

\bibitem{Vagnozzi:2017ilo}
S.~Vagnozzi,
``Recovering a MOND-like acceleration law in mimetic gravity,''
Class. Quant. Grav. \textbf{34}, no.18, 185006 (2017).


\bibitem{Mignani:1997a}
R.~Mignani, E.~Pessa and G.~Resconi, 
``Non-Conservative Gravitational Equations,''
Gen. Rel. Grav. \textbf{29}, 1049-1073 (1997).

\bibitem{Fabris:2017msx}
J.~C.~Fabris, H.~Velten, T.~R.~P.~Caram\^es, M.~J.~Lazo and G.~S.~F.~Frederico,
``Cosmology from a new nonconservative gravity,''
Int. J. Mod. Phys. D \textbf{27}, no.06, 1841006 (2018).


\bibitem{Gradshteyn:2015}
Gradshteyn,~I.~S.,
Ryzhik,~I.~M.,
Geronimus,~Y.~V., 
Tseytlin,~M.~Y., 
Jeffrey,~A.,
Zwillinger,~D.,
Moll, V.~H. (Eds.)
\textit{Table of Integrals, Series, and Products},
Academic Press (2015).

\bibitem{Pod:2003}
I.~Podlubny,
``Shadows on the Walls: Geometric Interpretation of Fractional Integration,''  
J. Online Math. Appl. \textbf{7} 1664 (2003).

\end{thebibliography}
\end{document}